\documentclass{PoS}

\title{Far-infrared-radio relation in cluster galaxies at intermediate redshift}

\ShortTitle{The far-infrared-radio relation in MS0451-03}

\author{\speaker{Solohery M. Randriamampandry}\\ 
        Astrophysics \& Cosmology Research Unit (ACRU), School of Maths, Statistics \& Computer Science, University of KwaZulu-Natal, Durban 4041, South Africa\\
        E-mail: \email{soloherymampionona@gmail.com}}

\abstract{The radio luminosities at 1.4 GHz is tightly correlated with the far-infrared luminosities for various galaxy types (e.g. \cite{van der Kruit 1971, Helou et al. 1985, Condon et al. 1991}) over a wide range of redshift (see e.g. \cite{Garrett 2002, Appleton et al. 2004, Sargent et al. 2010, Jarvis et al. 2010, Ivison et al. 2010}). The relationship is widely believed to be driven by the internal star formation activity. Radio emission from these galaxies are predominantly produced from the synchrotron emission of cosmic-ray electrons accelerated in supernova shocks. The infrared emission is due to ultraviolet light from young massive stars that is absorbed and re-radiated by dust \cite{Condon 1992}. A correlation is found also in local clusters but cluster galaxies appears to have excess radio emission relative to the amount of far-infrared emission \cite{MillerOwen 2001, ReddyYun 2004, Murphy et al. 2009}. In this work, we measure the far-infrared-radio relationship in a massive cluster to test how this relationship changes at intermediate z between the field and a high-density cluster environment.}

\FullConference{EXTRA-RADSUR2015 (*)\\
		20--23 October 2015\\
		Bologna, Italy

                \bigskip
                \hrule
                \bigskip

                \textnormal{(*) This conference has been organized
                  with the support of the Ministry of Foreign Affairs
                  and International Cooperation, Directorate General
                  for the Country Promotion (Bilateral Grant Agreement
                  ZA14GR02 - Mapping the Universe on the Pathway to
                  SKA)}
}

\begin{document}

\section{Motivation}
The main motivation in conducting this work consists of: (i) a number of work in the local Universe have reported that cluster galaxies appears to have excess radio emission relative to the amount of far-infrared emission \cite{MillerOwen 2001, ReddyYun 2004, Murphy et al. 2009}, (ii) so far, there is no investigations for the far-infrared-radio relation in clusters at higher z, (iii) thus this work aims to explore, for the first time, deviations or lack thereof of the far-infrared-radio relation in intermediate redshift cluster galaxies sample. In this contribution, we present results of studies performed for a massive galaxy cluster MS0451.6-0305 (hereafter, MS0451-03) at z$\sim$0.538. 

\section{Observations \& Method}
\begin{figure}
\begin{center}
\includegraphics[width=.46\textwidth]{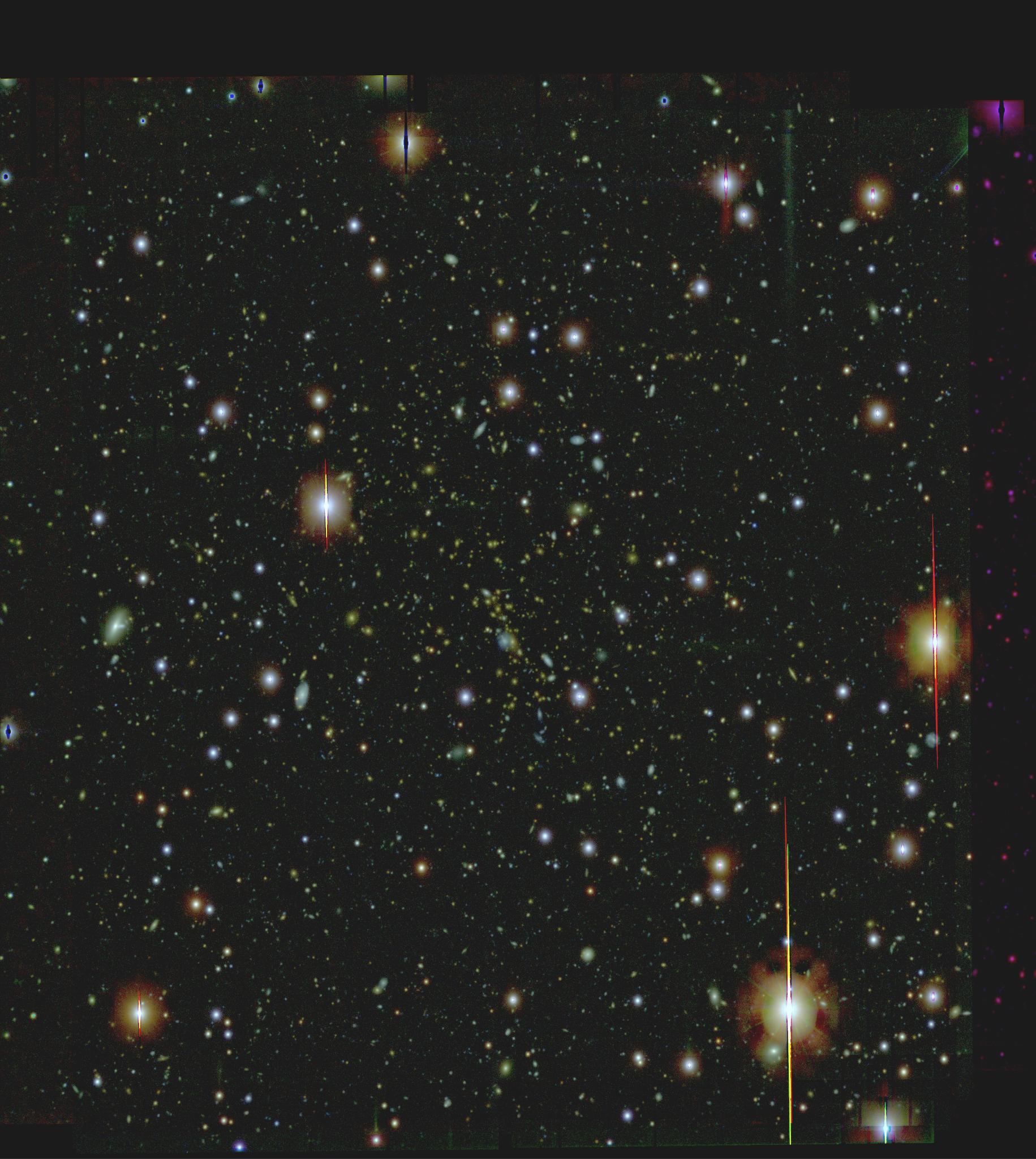}
\caption{Galaxy Cluster MS 0451.6-0305 at z=0.538.}
\label{fig:ms0451color}
\end{center}
\end{figure}

The multi-wavelength data used in this work consists of: (i) optical spectroscopic redshift from Keck \cite{Crawford et al. 2011} and from the literature (an optical image of cluster MS0451-03 is displayed in Figure \ref{fig:ms0451color}), (ii) VLA radio continuum observations at 1.4 GHz,  and (iii) Spitzer MIPS and IRAC super mosaic imaging.  

Radio data reduction and analysis were entirely carried out using the NRAO AIPS package. The L$_{\rm 1.4GHz}$ were derived following the method of \cite{Morrison et al. 2003}. The IR photometry and data analysis were done using Spitzer MOPEX APEX software. The L$_{\rm IR}$ were derived following the method of \cite{Rieke et al. 2009}.

\section{Results \& Summary}
\begin{figure}
\begin{center}
\includegraphics[width=.6\textwidth]{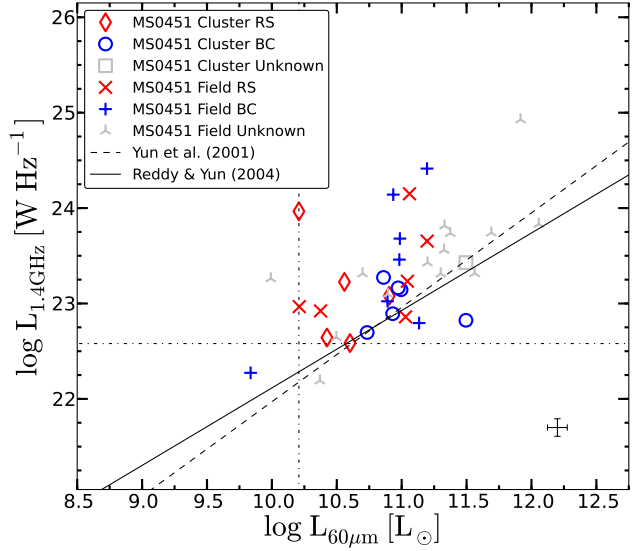}
\caption{The rest frame radio luminosity at 1.4 GHz (L$_{\rm 1.4GHz}$) and the FIR luminosity (L$_{\rm 60\mu m}$) (see \cite{Randriamampandry et al. 2015}).}
\label{fig:Fig}
\end{center}
\end{figure}

We present our results in Figure \ref{fig:Fig} (for more details see \cite{Randriamampandry et al. 2015}). The relationship between the rest frame radio luminosity at 1.4 GHz (L$_{\rm 1.4GHz}$) and the FIR luminosity (L$_{\rm 60\mu m}$) is shown. The solid line indicates the formal linear least-square fit of the cluster galaxies (see \cite{ReddyYun 2004}) while the field relation (see \cite{Yun et al. 2001}) is drawn using the dashed line.

In summary, we have constructed the far-infrared-radio relation for all sources with spectroscopic redshift using the Spitzer and VLA observations. We have found a correlation between the two parameters for the cluster star forming galaxies. We have measured q$_{\rm FIR}$-value of 1.80$\pm$0.15 with a dispersion of 0.53 for cluster galaxies which is in broad agreement, within uncertainties, with those for low z clusters (see \cite{Randriamampandry et al. 2015}).

\section{Future Work}
For further work, we: (i) shall expand the present sample into larger samples drawn from redMaPPer/ACT cluster sample in the SDSS Stripe 82 region and study the properties of star formation and AGN activity in clusters; (ii) shall use upcoming radio facilities such as MeerKAT/SKA which will enable us to further expand our sample to cover even more massive clusters at higher z to probe low luminosity star forming galaxies.

\section{Acknowledgments}
The financial assistance of the South African SKA Project (SKA SA) towards this research is hereby acknowledged. Opinions expressed and conclusions arrived at are those of the authors and are not necessarily to be attributed to the SKA SA (www.ska.ac.za).  SMR acknowledges the support of the Ministry of Foreign Affairs and International Co-operation, Directorate General for the Country Promotion (Bilateral Grant Agreement ZA14GR02- Mapping the Universe on the Pathway to SKA)

\end{document}